\documentclass[12pt]{article}
\usepackage{epsfig}
\usepackage{graphicx}
\DeclareGraphicsExtensions{.ps,.pdf,.png,.jpg}
\usepackage{amsmath}
\usepackage{amssymb}

\advance\hoffset by -7mm   
\makeatletter 
\@addtoreset{equation}{section}  
\makeatother 
 
\def\ii{\'\i}

\def\ftoday{{\sl {Le \number\day \space\ifcase\month 
\or janvier\or f\'evrier\or mars\or avril\or mai
\or juin\or juillet\or ao\^ut\or septembre\or octobre
\or novembre \or d\'ecembre\fi\space \number\year}}}    
\def\ptoday{{\sl {\number\day \space de\space \ifcase\month 
\or janeiro\or fevereiro\or mar{\c c}o\or abril\or maio
\or junho\or julho\or agosto\or setembro\or outubro
\or novembro \or dezembro\fi\space de\space \number\year}}}    
\def\gtoday{{\sl {Den \number\day. \ifcase\month 
\or Januar\or Februar\or M\"arz\or April\or Mai
\or Juni\or Juli\or August\or September\or Oktober
\or November \or Dezember\fi\space \number\year}}}    
\def\today{{\sl {\ifcase\month
\or January\or February\or March\or April\or May
\or June\or July\or August\or September\or October
\or November \or December\fi \space\number\day,\space 
                                            \number\year}}}
\newcommand{\journal}[4]{{\em #1~}#2\,(#3)\,#4}
\newcommand{\aihp}{\journal {Ann. Inst. Henri Poincar\'e}}

\newcommand{\ijmp}{\journal {Int. J. Mod. Phys.}}

\newcommand{\pr}{\journal {Phys. Rev.}}

\newcommand{\jmp}{\journal {J. Math. Phys.}}

\newcommand{\cqg}{\journal {Class. Quantum Grav.}}

\newcommand{\np}{\journal {Nucl. Phys.}}
\newcommand{\pl}{\journal {Phys. Lett.}}

\newcommand{\prep}{\journal {Phys. Rep.}}

\renewcommand{\a}{\alpha}
\renewcommand{\b}{\beta}
\newcommand{\g}{\gamma}           \newcommand{\GA}{\Gamma}
\renewcommand{\d}{\delta}         
\newcommand{\e}{\varepsilon}

\newcommand{\la}{\lambda}        
\newcommand{\m}{\mu}
\newcommand{\n}{\nu}
\newcommand{\om}{\omega}         
\newcommand{\p}{\psi}             
\newcommand{\s}{\sigma}           \renewcommand{\S}{\Sigma}
\newcommand{\f}{{\phi}}           \newcommand{\F}{{\Phi}}
\newcommand{\vf}{{\varphi}}

\newcommand{\CC}{{\cal C}}

\newcommand{\HH}{{\cal H}}

\newcommand{\LL}{{\cal L}}
\newcommand{\MM}{{\cal M}}

\newcommand{\esp}{\\[3mm]}

\newcommand{\sla}{\raise.15ex\hbox{$/$}\kern -.57em} 
\newcommand{\Sla}{\raise.15ex\hbox{$/$}\kern -.70em}
\def\h{\hbar}

\newcommand{\lp}{\left(}\newcommand{\rp}{\right)}

\newcommand{\complex}{{\kern .1em {\raise .47ex
\hbox {$\scriptscriptstyle |$}}
    \kern -.4em {\rm C}}}
\newcommand{\real}{{{\rm I} \kern -.19em {\rm R}}}
\newcommand{\rational}{{\kern .1em {\raise .47ex
\hbox{$\scripscriptstyle |$}}
    \kern -.35em {\rm Q}}}
\renewcommand{\natural}{{\vrule height 1.6ex width
.05em depth 0ex \kern -.35em {\rm N}}}

\newcommand{\half}{\frac{1}{2}}
\newcommand{\pa}{\partial}

\newcommand{\dfud}[2]{{\displaystyle{\frac{\delta #1}{\delta #2}}}}
\renewcommand{\dfrac}[2]{{\displaystyle{\frac{#1}{#2}}}}
   
\newcommand{\dint}{{\displaystyle{\int}\!\!\!}}

\newcommand{\ie}{{{\em i.e.},\ }}

\newcommand{\twiddle}{\lower.9ex\rlap{$\kern -.1em\scriptstyle\sim$}}

\newcommand{\ket}[1]{\left| {#1}\right\rangle}

\newcommand{\vev}[1]{\langle {#1}\rangle}

\newcommand{\equ}[1]{(\ref{#1})}
\newcommand{\eq}{\begin{equation}}
\newcommand{\eqn}[1]{\label{#1}\end{equation}}
\newcommand{\eea}{\end{eqnarray}}
\newcommand{\eqa}{\begin{eqnarray}}
\newcommand{\eqan}[1]{\label{#1}\end{eqnarray}}
\newcommand{\ba}{\begin{array}}
\newcommand{\ea}{\end{array}}
\newcommand{\eqac}{\begin{equation}\begin{array}{rcl}}
\newcommand{\eqacn}[1]{\end{array}\label{#1}\end{equation}}

\newcommand{\bz}{\begin{enumerate}}
\newcommand{\ez}{\end{enumerate}}

\newcommand{\bsx}{{\boldsymbol{x}}}
\newcommand{\bsy}{{\boldsymbol{y}}}



\newcommand{\bfi}{{\bar{\f}}}
\newcommand{\onsh}{\stackrel{*}{=}}
\newcommand{\tF}{{\tilde F}}
\newcommand{\tB}{{\tilde B}}
\newcommand{\te}{{\tilde e}}

\begin{document}


\title{Loop quantization of a 3D Abelian BF model \\ with $\s$-model matter}
\author{Diego C. M. Mendon{\c{c}}a\footnote{Work supported
   in part by the Coordena\c c\~ao de Aperfei\c coamento de Pessoal de N\' ivel Superior -- CAPES (Brazil).}\\
{\small Departamento de F\ii sica, Universidade Federal do Esp\'{\i}rito Santo (UFES)}\\
{\small Vit\'oria, ES, Brazil.}\esp
and  Olivier Piguet\footnote{Work supported
   in part by the Conselho Nacional de Desenvolvimento Cient\'{\i}fico e
   Tecnol\'{o}gico -- CNPq (Brazil).}\\
   {\small Departamento de F\ii sica, 
Universidade Federal de Vi\c cosa -- UFV, Vi\c cosa, MG, Brazil}}
\date{\today}
\maketitle


\begin{center} 

\vspace{-5mm}

{\small\tt 
E-mails: diegomendonca@gmail.com, opiguet@yahoo.com}
\end{center}

\vspace{5mm}

\begin{abstract}

The main goal of this work is to explore the symmetries and develop the dynamics associated to a 3D Abelian BF model coupled to scalar fields submitted to a sigma model like constraint, at the classical and quantum levels. We adapt to the present model the techniques of Loop Quantum Gravity, construct its physical Hilbert space and its observables.

\end{abstract}

\section{Introduction}

The now quasi-hundred years old General Relativity as a theory of gravitation, despite of its tremendous successes in accounting for predicting phenomena, still lacks of a quantum version. Previous perturbative attempts have shown the non-renormalizability of the theory~\cite{thooft}, whereas the pioneering non-perturbative approach of Wheeler and DeWitt~\cite{WdW} had its successes concentrated in reduced ``minisuperspace''	models dedicated to Cosmology.
However, very important progresses have been made in the last decades, especially in the framework of Loop Quantum Gravity (LQG)~\cite{general-ref}, based on the canonical Hamiltonian approach of Dirac and Bergman~\cite{dirac} applied to the Ashtekar-Barbero~\cite{ashtekar,barbero-immirzi} parametrization of the theory. General Relativity, as a background independent theory -- in the sense that no background geometry is given {\it a priori}, geometry being dynamical -- is a fully constrained theory, its Hamiltonian being merely  a sum of constraints generating the gauge invariances of the theory.
The LQG program entails the difficult task of implementing the constraints of the theory as quantum operators in some predefined kinematic Hilbert space, and to solve them, thus leaving as a subspace the physical Hilbert space in which act the self-adjoint operators representing the observables of the theory. Some of the constraints have been resolved, but a last one, the so-called scalar constraint. The latter has resisted up to now a complete solution, the most popular approach being that of ``spin foams''~\cite{zapa1,perez1}.

By contrast, the lower-dimensional gravitation theories are much more easy to handle, since they can be described as topological gauge theories, when not coupled to matter~\cite{Achucarro-etc,carlip-gegenberg,carlip-gegenberg-mann,freidel-mann-popescu,nouiperez}. Coupling them to matter however lets them loose their topological character, excepted in some special cases, where a complete and rather simple loop quantization can be achieved~\cite{dis_diego,Diego-Olivier}.

The purpose of this paper is to present the loop quantization of a topological theory of the BF type~\cite{baez2} with the Abelian group U(1)
as a gauge group. The BF fields are coupled to a complex scalar ``matter'' field subject to a $\s$-model type of constraint. It turns out that the topological nature of the theory persists in the sense that no local degrees of freedom are present. The physical Hilbert space is constructed with a non-trivial result if the topology of space is non-trivial. Spaces with point-like singularities are considered, in which cases global observables are explicitly constructed. A non-Abelian version is presently under study~\cite{Diego-Olivier}.

The model and its gauge invariances is presented in Section 2, its classical analysis is done in Section 3 together with the separation of the first and second class constraints and the definition of the Dirac brackets, and the quantization is presented in Section 4. Brief conclusions are given at last.

\section{Formulation of the model}
\subsection{The gauge invariances and the action}

The field content of the model is a U(1) connection form 
$A=A_\m(x)dx^\m$, a ``B'' form $B=B_\m(x)dx^\m$,
a complex scalar field $\f(x)$ and a 3-form field 
$e=\frac{1}{3!}e_{\m\n\rho} dx^\m dx^\n dx^\rho$, 
transforming 
as\footnote{Wedge symbols $\wedge$ are not written explicitly. Space-time indices $\m,\n\cdots$ take the values 0,1,2; later on, space indices will be denoted by the letters $a,b\cdots$ taking the values 1,2. $A_\m$ and $\theta$ are taken as imaginary.}
\eq
A_\m'=A_\m' + g^{-1}\pa_\m g = A_\m  +  \pa_\m\theta \,,
\quad B'=B\,,\quad \f'=g\f\,, \quad \bfi'=g^{-1}\bfi\,,
\quad e'=e\,,
\eqn{transf-g}
under U(1) gauge transformations\footnote{$A_\m$ and $\theta$ are taken as purely imaginary;
$B_\m$ is real.} $g(x)=\exp\theta(x)$ 

One introduces also the topological type gauge transformations
\eq
A'=A\,,\quad B' = B + d\p\,,\quad \f'=\f\,,\quad
e' = e -d\p F\,,
\eqn{transf-psi}
where $F=dA$ and the scalar $\p(x)$ is the transformation parameter.
These transformations coincide with the usual topological type transformations of the $BF$ model, in the absence of the fields $\f$ and $e$.

The most general action invariant under the sole gauge 
transformations \equ{transf-g} can be written as
\[
S_{\rm general} = \dint_{\MM^3} \lp  K(\bar\f\f) BF 
+ \la(\bar\f\f)B D\bar\f D\f + e[\m(\bar\f\f)-R] \rp\,,
\]
where $ K,\,\la$ and $\m$ are arbitrary functions of $\bar\f\f$,
and $D$ denotes the covariant  derivative(covariant with respect to the gauge transformation \equ{transf-g}):
\[
D\f = (d-A)\f\,,\quad D\bar\f = (d+A)\bar\f\,.
\]
The integration is performed over some 3D differential 
manifold $\MM^3$. The action is obviously invariant under the diffeomorphisms of $\MM^3$.

The parameter $R$ can be taken equal to 1 through a renormalization of the field $e$, and one easily shows that one can reduce the function $\m(\bar\f\f)$ to the form $\m=\bar\f\f$ through a suitable field redefinition $\phi\to\phi'(\phi,\bar\f)$, 
$\bar\phi\to\bar\phi'(\phi,\bar\f)$, compatible with the gauge transformation \equ{transf-g}. 
Imposing now the topological gauge invariance $\equ{transf-psi}$ fixes the function $\la$ to the constant value 1.  The  action then reads
\[
 \dint_{\MM^3} \lp  K(\bar\f\f) BF 
+ B D\bar\f D\f + e(\bar\f\f-1) \rp\,.
\]
The resulting field equation $\bar\f\f=1$ implies that the function $K$ can be replaced by a constant, which in turn can be reabsorbed through a renormalization of the field $B$. The final action is then
\eq
S= \dint_{\MM^3} \lp  BF + B D\bar\f D\f + e(\bar\f\f-1) \rp\,.
\eqn{action}

One recognizes in \equ{action} a $BF$ action coupled with scalar fields and a Lagrange
 multiplier field $e$ assuring the $\s$-model type constraint
  $\bar\f\f=1$. 

It turns out that the action \equ{action} has a third gauge invariance: It is invariant under the following local transformations, of parameter $\eta(x)$:
\eq
A' = A + d\eta\,,\quad B'=B\,,\quad, \f'=\f\,,\quad
e' = e + d\eta\, dB\,.
\eqn{transf-eta}

In order to check the invariances of the action (up to boundary terms), as well as for all the manipulations involving partial integrations, it is useful to remember that the covariant derivative $D$, defined by
$DX=dX-qAX$ where $q$ is the U(1) charge of the field X, obeys the Leibniz rule. The respective U(1) charges of the basic fields $A,B,\f,\bar\f$ and $e$ are $0,0, 1,-1$ and 0. Let us also note the useful identity
\[
D\bar\f D\f = d\bar\f d\f + A\,d(\bar\f\f)\,.
\]
The field equations read
\[\ba{ll}
\dfud{S}{B} = F +D\bar\f D\f 
                    \stackrel{*}{=} 0\,,\quad
&\dfud{S}{A} = { }dB -Bd(\bar\f\f) \onsh 0  \,,\esp
\dfud{S}{\bar\f} = BF\f  -dBD\f + e\f \onsh 0\,,\quad
&\dfud{S}{\f}=  BF\bar\f -dBD\bar\f + e\bar\f\onsh 0\,,\esp
\dfud{S}{e} = \bar\f\f - 1 \onsh 0\,,
\ea\]
where the symbol $\onsh$ means ``on shell'' equality, i.e., 
``equations of motion being fulfilled''. 
The last equation is equivalent to
\eq
\f(x) \onsh e^{i\vf(x)}\,,\quad \bfi(x) \onsh e^{-i\vf(x)}\,,\quad
\vf\mbox{ a real phase}\,.
\eqn{phi=phase}
This system of equations is equivalent to the simpler one:
\[
F \onsh 0\,,\quad dB \onsh 0\,,\quad e \onsh 0\,,
\quad \bar\f\f - 1 \onsh 0\,.
\]

\subsection{Diffeomorphism invariance}

In the present theory,  like in the topological theories of the Chern-Simons or $BF$ type, the invariance under the diffeomorphisms  is a consequence of the invariance under the gauge transformations \equ{transf-g}, \equ{transf-psi} and \equ{transf-eta}, up to field equations. 
Indeed, the diffeomorphisms being generated by the Lie derivative
$\LL_\xi=i_\xi d+di_\xi$ along an infinitesimal vector field $\xi$ when acting on forms\footnote{$i_\xi$ is the interior derivative, 
with $i_\xi dx^\m = \xi^\m$.}, 
one checks that
\[\ba{ll}
\LL_\xi A \onsh d(i_\xi A)\,,\quad &\LL_\xi B = d(i_\xi B)\,,\esp
\LL_\xi\f \onsh i(i_\xi d\vf)\f\,,&\quad
\LL_\xi\bfi \onsh - i(i_\xi d\vf)\bfi\,,\quad
\LL_\xi e \onsh 0\,,
\ea\]
where $\vf$ is the phase of the field $\f$ defined in \equ{phi=phase}. One sees that these infinitesimal diffeomorphisms are given, on-shell, by a combination of the three gauge invariances, with the respective 
field dependent infinitesimal parameters given by
\[
\theta = i(i_\xi d\vf)\,,\quad \psi =i_\xi B\,,\quad
\eta = i_\xi (A -i d\vf)\,.
\]

\section{Hamiltonian analysis and constraints}

We apply here the canonical formalism of Dirac~\cite{dirac} for systems with constraints.
Supposing that the space-time manifold  admits a ``time''~$\times$~``space''
foliation $\MM_3=\mathbb{R}\times\S$, where the space slice $\S$ is some two-dimensional manifold, we first rewrite the action as the time integral 
\[
S=\int dt\, L(A,\dot A,B,\dot B,\f,\dot \f,\bar\f,\dot{\bar\f},e,\dot e)
\]
of a Lagrangian function
\eq\ba{l}
L(A,\dot A,B,\dot B,\f,\dot \f,\bar\f,\dot{\bar\f},e,\dot e) = \esp
\ \dint_{\S} d^2x 
\lp 
 { }\tB^a \pa_tA_a - \tB^a D_a\bar\f\,\pa_t\f + \tB^a D_a\f\,\pa_t\bar\f
+ A_t \CC_1 + B_t\CC_2 + \te\CC_5\rp\,,
\ea\eqn{Lagrangian}
where
\eq\ba{l}
\CC_1 =  { }\pa_a\tB^a +\tB^a\pa_a(\bar\f\f)\esp
\CC_2 =  { }\tF + \e^{ab}D_a\bar\f D_b\f \esp
\CC_5 = \bar\f\f-1 \,,\esp
\tF = \half\e^{ab}F_{ab}\,,\quad\tB^a=\e^{ab}B_b\,,\quad
\te= \frac{1}{3!}\e^{\m\n\rho} e_{\m\n\rho}\,.
\ea\eqn{constraints}
Following the canonical procedure, we identify the conjugate momenta of each field $X$, $\Pi_X=\d{L}/\d{\dot X}$:
\eq\ba{l}
\Pi_{A_t}=0\,,\quad  \Pi_{B_t}=0\,, \quad\Pi_{\te}=0\,,
\Pi_{\f}=-\tB^a D_a\bar\f\,,\quad \Pi_{\bar\f}=\tB^a D_a\f\,,\esp
\Pi_{A_a}= \tB^a\,,\quad    \Pi_{B_a}=0\,,
\ea\eqn{momenta}
satisfying together with the $X$'s the equal time Poisson bracket relations
\[
\{X^\a(\bsx), \Pi_{X^\b}(\bsy)\} = \d^\a_\b \d^2(\bsx,\bsy)\,,
\quad \{X^\a(\bsx), X^\b(\bsy)\} = \{\Pi_{X^\a}(\bsx), \Pi_{X^\b}(\bsy)\} = 0\,,
\]
where the indices $\a,\b$ run over all components of all fields.
The Legendre  transform 
$H_{\rm c}= -L + \sum_\a\int d^2x\,\Pi_{X^\a} {\dot X}^\a$
yields the canonical Hamiltonian
\[
H_{\rm c}= -\int d^2x\lp A_t \CC_1 + B_t\CC_2 + \te \CC_5 \rp\,,
\]
with the $\CC$'s given in \equ{constraints}.

Noting that the velocities don't appear in any of the equations
(\ref{momenta})  for the momenta, we conclude that all of these equations  are (primary) constraints~\cite{dirac}. The equality sign must be replaced by the ``weak equality'' sign $\approx$, meaning that the constraints are solved at the end,  after all calculations involving Poisson brackets are done. We remark that the last two constraints in \equ{momenta} are second class, their brackets being non-zero:
$\{\Pi_{A_a}(\bsx)-\e^{ab}B_b(\bsx),\,\Pi_{B_c}(\bsy)\}$ $=$ $\e^{ab}\d^2(\bsx,\bsy)$. These constraints can be solved as strong equalities 
\eq
\Pi_{A_a}= \e^{ab}B_b=\tB^a\,,\quad \Pi_{B_a}=0\,,
\eqn{A-moment}
provided the Poisson Brackets are replaced by the corresponding Dirac 
brackets, which read
\[
\{A_a(\bsx),\,\tB^b(\bsy)\} = \d^b_a \d^2(\bsx,\bsy)\,,
\]
the other brackets being left unchanged. We use the same notation 
$\{\cdot,\cdot\}$ for these Dirac brackets.

We are left with the five constraints
\eq
\Pi_{A_t}\approx 0\,,\quad  \Pi_{B_t}\approx 0\,, 
\quad\Pi_{\te}\approx 0\,,
\eqn{constraint1}
and
\eq 
\CC_3(\bsx)=\Pi_{\f}+\tB^a D_a\bar\f \approx 0\,,\quad 
\CC_4(\bsx)=\Pi_{\bar\f}-\tB^a D_a\f \approx 0\,.
\eqn{constraint3}
The stability of the three constraints \equ{constraint1} under the Hamiltonian evolution requires the three secondary constraints
\eq
\CC_1(\bsx) \approx 0\,,\quad \CC_2(\bsx) \approx 0\,,\quad \CC_5(\bsx) \approx 0\,,\quad\,,
\eqn{sec-constraints}
with  $\CC_1$, $\CC_2$ and $\CC_5$ as
given in \equ{constraints}. It will turn out convenient to replace 
$\CC_1$ by the equivalent constraint:
\eq
\CC'_1(\bsx)\approx 0\,,\quad\mbox{with}\quad  
\CC'_1(\bsx) = \CC_1 -\f\CC_3 + \bfi\CC_4 
= \pa_a\tB^a -\f\Pi_\f +\bfi\Pi_\bfi \,.
\eqn{constraint-C1}
The constraints \equ{constraint1} can be put strongly to zero, 
the corresponding fields $A_t$, $B_t$ and $\te$ playing  now the roles of Lagrange multipliers $\la_1$, $\la_2$ and $\la_5$. Introducing also Lagrange multipliers fields for the primary constraints \equ{constraint3}, we define the total Hamiltonian as\eq
H_{\rm T} =\sum_{m=1}^5 \CC_m[\la_m]\,,
\eqn{tot-ham}
where we have defined the functionals 
\eq
\CC_m[\la_m]= \int d^2x\, \la_m(\bsx)\CC_m(\bsx)\,,
\eqn{functional-notation}
considering the Lagrangian multiplier 
fields $ \la_m$ as smooth test functions.

Since this Hamiltonian is entirely made of  constraints -- a characteristics of theories with general covariance  -- the stability of our five constraints $\CC_\a$, $\a=1,\cdots,5$ 
amounts to examine the matrix $M_{mn}(\bsx,\bsy)$ $\approx$ 
$\{\CC_m(x),\CC_n(y)\}$
of their Poisson brackets -- written up to constraints, hence the $\approx$ sign. Indeed, their stability condition reads
(summation convention is assumed)
\eq 
\dot\CC_m=\{\CC_m,H_{\rm T}\}=M_{mn}\la^n=0\,.
\eqn{stab-cond}
This provides a system of equations for the $\la$', which can be solved for
some of the  $\la$'s in terms of the remaining ones.
The matrix $M$ reads
\[
M=
\lp\ba{ccccc}
0&0&0&0&0\\
0&0&0&\f\CC_2&0\\
0&0&0&-(\f\CC_1'+\CC_3)&0\\
0&-\f\CC_2&(\f\CC_1'+\CC_3)&0&-\f\\
0&0&0&\f&0\
\ea\rp 
\d^2(\bsx-\bsy)\,,
\]
where we have substituted the constraint $\CC_3$ with the equivalent one
\[
\CC_3'=\f\,\CC_3-\bar\f\,\CC_4\,.
\]
One sees that the first three constraints, $\CC_1$, $\CC_2$ and $\CC'_3$ 
are first class, \ie their Poisson brackets with any other constraint 
are constraints: they generate three gauge invariances of the theory.
 The two last ones, namely $\CC_4$ and $\CC_5$ however are second class. 
 Indeed, denoting them by $\chi_p$ ($p=1,2$), their Poisson brackets 
 form the matrix $C_{pq}$ of non-vanishing determinant on 
 the constraint surface:
\[
C=\lp\ba{cc}0&-\f\\\f&0\ea\rp\,.
\]
These second class constraints may be written as strong equalities,
 provided the Poisson brackets are substituted by the 
 Dirac brackets~\cite{dirac}
\eq 
\{X,Y\}_{\rm D} =\{X,Y\} -\sum_{p,q} \{X,\chi_p\}(C^{-1})^{pq}\{\chi_q,Y\}\,.
\eqn{Dirac-brackets}
The second class constraints $\chi_p$ can be solved for $\bfi$ and $\Pi_{\bfi}$ in terms of the now independent fields $A_a$, $\tB^a$, $\f$ and $\Pi_\f$,
\[
\bfi=1/\f\,,\quad \Pi_{\bfi}=\tB^aD_a\f\,.
\] 
The independent fields obey the Dirac bracket relations
\[\ba{ll}
\{A_a(\bsx),\tB^b(\bsy)\}_{\rm D}= \d^b_a\,\d^2(\bsx-\bsy)\,,\quad&
\{\f(\bsx),\Pi_\f(\bsy)\}_{\rm D}= \d^2(\bsx-\bsy)\,,\esp
\{A_a(\bsx),\Pi_\f(\bsy)\}_{\rm D}= D_a\lp\dfrac{1}{\f}\rp\,\d^2(\bsx-\bsy)\,,\quad&
\{\tB^a(\bsx),\Pi_\f(\bsy)\}_{\rm D}= -\tB^a\dfrac{1}{\f}\d^2(\bsx-\bsy)\,,\esp
\mbox{(other brackets vanishing)}\,.
\ea\]
This system can be diagonalized through the redefinition
\eq
\Pi=\Pi_\f-\tB^a D_a\lp\dfrac{1}{\f}\rp\,,
\eqn{sol-2ndclass-constr}
with the result:
\eq\ba{l}
\{A_a(\bsx),\tB^b(\bsy)\}_{\rm D}= \d^b_a\,\d^2(\bsx-\bsy)\,,\quad
\{\f(\bsx),\Pi(\bsy)\}_{\rm D}= \d^2(\bsx-\bsy)\,,\esp
\mbox{(other brackets vanishing)}\,.
\ea\eqn{Dirac-brackets'}
Finally, the remaining three constraints read, taking \equ{sol-2ndclass-constr} into account:
\eq
\CC_1=\pa_a \tB^a\,,\quad \CC_2 = \tF ,\quad \CC_3 = \F\Pi\,.
\eqn{class-constr}
They are first  class (their Dirac brackets are indeed zero), and generate the three
 gauge invariances defined by $\d_i X$ $=$ $\{X,\CC_i[\epsilon_i]\}$ ($i=1,2,3$)
  using the functional notation \equ{functional-notation}:
\eq\ba{lll}
\d_1 A_a =  -\pa_a\epsilon_1 \,,\quad&
\d_2 A_a  = 0\,,\quad&
\d_3 A_a  = 0\,,\quad\esp
\d_1 \tB^a =0 \,,\quad&
\d_2 \tB^a =  -\epsilon^{ab}\pa_b\epsilon_2\,,\quad&
\d_3 \tB^a = 0\,,\quad\esp
\d_1 \f = 0 \,,\quad&
\d_2 \f = 0\,,\quad&
\d_3 \f = \epsilon_3\f\,,\quad\esp
\d_1 \Pi = 0 \,,\quad&
\d_2 \Pi = 0\,,\quad&
\d_3 \Pi = -\epsilon_3\Pi\,.
\ea\eqn{g-invariances}
We see that the U(1) gauge invariance is split in two invariances generated
 by  $\CC_1$ and $\CC_3$, corresponding to the invariances \equ{transf-g}
  and \equ{transf-eta} of the Lagrangian formalism. The invariance 
  generated by $\CC_2$ corresponds to the topological type invariance
   \equ{transf-psi}.

\section{Quantization}
\subsection{Kinematical Hilbert space}   

The constraints $\CC_1$ and $\CC_3$ will be solved at the quantum level 
in this  Section, whereas the last one, $\CC_2$, is left for 
the next Section. Following the lines of Loop Quantum 
Gravity~\cite{general-ref}, we shall construct a kinematical Hilbert space 
$\HH_{\rm kin}$ whose vectors $\ket{\Psi}$ are subjected to the constraints
$\CC_1$ and $\CC_3$ in the form
$\hat\CC_1\ket{\Psi}=0$ and $\hat\CC_3\ket{\Psi}=0$, where $\hat\CC_i$ are operators representing the classical $\CC_i$.
Choosing the fields $A_a$ and $\f$ as configuration 
space coordinates, our task will be to define wave 
functionals\footnote{We use the ``bra'' and ``ket'' Dirac notation, with 
$\vev{A,\f|\Psi}$ = $\Psi[A,\f]$.} 
$\Psi[A,\f]$ and the scalar product $\vev{\Psi|\Psi'}$. 
The fields are now promoted to operators $\hat A_a$, $\hat\f$, $\hat B^a$ 
and $\hat\Pi$ obeying the canonical commutation relations corresponding 
to the classical Dirac brackets\equ{Dirac-brackets'}:
\eq\ba{l}
[\hat A_a(\bsx),\hat B^b(\bsy)]= i\h\d^b_a\,\d^2(\bsx-\bsy)\,,\quad
[\hat\f(\bsx),\hat\Pi(\bsy)]= i\h\d^2(\bsx-\bsy)\,,\esp
\mbox{(other brackets vanishing)}\,.
\ea\eqn{commutators}
$\hat A$ and $\hat\f$ act multiplicatively, $\hat B$ and $\hat\Pi$ 
as functional derivatives:
\[
\hat B^a(\bsx)\Psi[A,\f] = -i\h \dfud{\Psi[A,\f]}{A_a(\bsx)}\,,\quad
\hat\Pi(\bsx)\Psi[A,\f] = -i\h \dfud{\Psi[A,\f]}{\f(\bsx)}
\]
 Everything up to now is purely formal since we have still 
no proper Hilbert space. But we can already solve the constraint 
$\hat\CC_3(\bsx)\Psi[\f,A]$ = 
$-i\h\hat\f\,\d\Psi[\f,A]/\d\f(\bsx)$ $=0$: the wave functional only depends on $A$, $\Psi$ = $\Psi[A]$.

In order to construct a scalar product defined by an appropriate 
integration measure in configuration space, we first restrict the 
space of wave functionals to  the set of functions of finite numbers of holonomies of the connection $A$  -- the ``cylindrical functions''. If $\g$ is an orientated curve in $\S$ (a ``link''), the holonomy of $A$ on $\g$ is defined as the exponentiated line integral
\eq
h_\g[A] = \exp{\int_\g A}\,.
\eqn{holonomy}
Given a ``graph'', \ie a finite set $\GA=\{\g_1,\cdots,.\g_N\}$ of links, a ``cylindrical function'' $\Psi_{\GA,\psi}[A]$ is function $\p$ of the holonomies of $\GA$:
\[
\Psi_{\GA,\psi}[A]=\p(h_{\g_1}[A],\cdots,h_{\g_N}[A])\,.
\]
The cylindrical functions associated to all graphs on $\S$ form 
the vectorial space Cyl, in which we can define a sesquilinear scalar product 
using the Haar measure $d\m(g)$ of the gauge group. For U(1), the (normalized)
 measure is given by $\frac{1}{2\pi}\int d\theta f(g(\theta))$ 
 for $g$ parametrized as $g(\theta)=\exp(i\theta)$. 
 First, for two cylindrical functions defined on the same graph: 
\[
\vev{\GA,\p|\GA,\p'} = \int_{G^{\otimes N}}\int d\mu(g_1)\,\cdots d\mu(g_N)
(\p(g_1,\cdots,g_N))^*\p'(g_1,\cdots,g_N)\,.
\]
Next, for two cylindrical functions corresponding to two different graph $\GA$ and $\GA'$, 
one defines 
\[
\vev{\GA,\p|\GA',\p'} = 
\int_{G^{\otimes N}}\int d\mu(g_1)\,\cdots d\mu(g_{\hat N})
(\p(g_1,\cdots,g_N))^*\p'(g_1,\cdots,g_{N'})\,.
\]
where  $\hat\GA$ is the union graph $\GA\cup\GA'$ consisting of $\hat N\leq(N+N')$ links.

With this scalar product in hands we dispose of a norm so one can define a Hilbert space $\HH_{\rm Cyl}$ through the Cauchy completion of Cyl. 

An orthonormal basis of $\HH_{\rm Cyl}$ may be defined using the Peter-Weyl theorem -- which in the Abelian U(1) case is nothing but the Fourier series theorem. Basis elements are the cylindrical functions
\eq\ba{c}
\Psi_{\GA,\vec n}[A]=\chi_{n_1}\lp h_{\g_1}[A]\rp\cdots
\chi_{n_N}\lp h_{\g_N}[A]\rp\,,\esp
\mbox{where }\vec n=(n_1,\cdots n_N)\,,\quad n_k\in \mathbb{Z}\,,\quad n_k\not=0\,,
\ea\eqn{basis-Cyl}
and  
$\chi_n(g)$ is the character of the irreducible unitary representation of 
``charge'' $n\in \mathbb{Z}$. In the parametrization $g=\exp(i\theta)$, 
$\chi_n(g)$ = $\exp(in\theta)$. The orthonormality condition
\[
\vev{\GA,\vec n|\GA',{\vec n}'}\,,
\]
is an obvious consequence of the theory of Fourier series.
The prescription of non-vanishing charges $n_k$ avoids an over-counting of the basis vectors which would otherwise occur since a graph with a zero charge link would give the same function as the graph with this link omitted. Therefore, the basis must be completed with the zero charge function $\Psi_\emptyset$ corresponding to the empty set $\emptyset$. These basis vectors 
$\ket{\GA,\vec n}$ will be called 
``charge networks'' in analogy with the spin networks of 
Loop Quantum Gravity~\cite{general-ref}. 
A particular consequence of these definitions is that vectors corresponding to different graph are orthogonal, and thus the Hilbert space $\HH_{\rm Cyl}$ is 
the infinite direct sum of spaces $\HH_{\rm Cyl,\GA}$, each of them 
being associated to a single graph $\GA$. This sum being performed over the 
non-countable set of all graphs, $\HH_{\rm Cyl}$ is a non-separable 
Hilbert space.

\begin{figure}[h]
\begin{center}
\includegraphics[scale=.55]{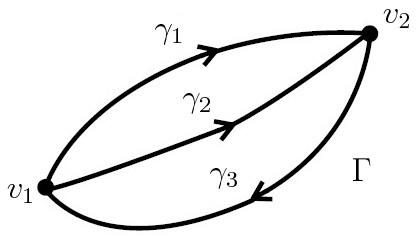}
\caption{Closed graph $\GA$ with three links and two vertices.
Link $\g_k$ carrying a charge $n_k$, charge conservation 
at vertex $v_1$ (or $v_2$)  amounts to $-n_1-n_2+n_3=0$.}\label{fig1}
\end{center}
\end{figure}

Let us now turn to the constraint $\CC_1$ in \equ{class-constr}, which corresponds to the invariance under the U(1) gauge transformations $\d_1$ of \equ{g-invariances}. It will be fulfilled by demanding the gauge invariance of the basis cylindrical functions \equ{basis-Cyl}. 
Under a gauge transformation $A'_a$ = $A'_a+\pa_a\om$, the
holonomy \equ{holonomy} transforms as 
\[
h_\g[A]'=h_\g[A]\exp(\om(\bsx_{\rm f})-\om(\bsx_{\rm i}))\,,
\]
where $\bsx_{\rm i}$ and $\bsx_{\rm f}$ are the coordinates of the initial and end points of the link $\g$, respectively. Thus gauge invariance of a charge network functional $\Psi_{\GA,\vec n}$ follows from the requirement of a ``charge conservation law'', \ie the sum of charges entering a vertex of $\GA$ (point of intersection of links) must be zero, with the convention that the charge entering a vertex is positive if the vertex lies at the end of the link, and negative if it lies at the beginning. This requires in particular that the graphs must be closed since no zero-charge links are allowed.
An example is depicted in Fig.~\ref{fig1}.

The vectors of $\HH_{\rm Cyl}$ obeying the condition of gauge invariance span the non-separable ``kinematical'' Hilbert space $\HH_{\rm kin}$ $\subset$ $\HH_{\rm Cyl}$.

\subsection{Physical Hilbert space}

The last constraint to be imposed is the curvature constraint 
$\CC_2$ in \equ{class-constr}, whose quantum expression is
$\hat{F}\ket{\Psi}=0$. Its  general solution is given by a wave 
functional $\Psi[A]$ whose argument A is a connection with null curvature. It is sufficient to impose this condition on the basis vectors of  $\HH_{\rm kin}$ (charge networks), 
which will select the basis  of the physical Hilbert space 
$\HH_{\rm phys}$ $\subset$ $\HH_{\rm kin}$. 

The condition of null curvature means that, {\it locally}, there exists a scalar function $\vf$ such that 
\eq
A_a=\pa_a\vf\,.
\eqn{zero-curv}
 The rest of the discussion depends on the topology of the space sheet $\S$.

Let us begin with the case where the topology of $\S$ is that of $\mathbb{R}^2$.
Then \equ{zero-curv} holds globally, with the result that the 
holonomy  associated to any link $\g$ with initial and final end points 
$\bsx_{\rm i}$ and $\bsx_{\rm f}$ takes the form
\[
h_\g[A]=\exp(\vf(\bsx_{\rm f})-\vf(\bsx_{\rm i}))\,.
\]
Together with the fact that that the graph associated to any charge network
$\ket{\GA,\vec n}$ is closed and that the charge conservation 
condition must hold at each vertex, one easily sees that  
its wave functional $\Psi_{\GA,\vec n}$ is equal to 1. In other words, 
the graph $\GA$ shrinks to a single point, and we are left with the 
sole vector $\ket{\emptyset}$. The physical Hilbert space is reduced to
a trivial 1-dimensional space.

\begin{figure}[h]
\begin{center}
\includegraphics[scale=.8]{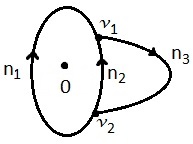} \quad
\includegraphics[scale=.75]{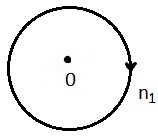}
\caption{Two charge network graphs with the singular point O ``inside''.}\label{fig2}
\end{center}
\end{figure}

The next case is that with the topology of $\mathbb{R}^2\backslash\{O\}$, 
the 2-dimensional plane with one point $O$ suppressed. There are now two classes of closed graphs, those with $O$ inside and those with $O$ outside. Two examples of the former class are shown in Fig.~\ref{fig2}.

Applying the charge conservation condition as in the previous case shows 
that any charge network graph with the point $O$ ``outside'' 
reduces to a point with the resulting wave functional equal to 1, defining the empty state described by the vector  $\ket{\emptyset}$.
On the other hand, any charge network graph with the point $O$ ``inside'' 
is equivalent to a single loop $\g$ with $O$ inside, with the resulting 
wave functional equal to a unimodular complex number: 
\[
\vev{A|n}= \Psi_{n}[A]=\exp(inQ)\,,
\] 
with $n\in\mathbb{Z}$ the charge of the loop.
The value of the ``flux'' $Q$, given by
\[
\exp(iQ)= h_C[A]\,,
\]
where $C$ is a closed positively oriented loop around the singular point $O$, 
 is independent of the form and size of 
the loop, and the value of $n$ is computed using the charge 
conservation condition.
 Fig.~\ref{fig2} shows an example of two such equivalent graphs.
 The basis of the physical Hilbert space $\HH_{\rm phys}$ then consists of the vectors
 $\ket{n}$, $n\in\mathbb{Z}$, with $\vev{n|n'}=\d_{nn'}$. For $n=0$, one has 
 $\ket{0}$ = $\ket{\emptyset}$, corresponding to the former class of graphs.
 One notes that the integer number $n$ can be interpreted as a winding number of the loop: to wind $n$ times around the singular point with charge 1, or to wind 1 time with charge $n$ yield the same wave functional.

The generalization to a plane with $N$ singular points, 
$\mathbb{R}^2\backslash\{O_1,\cdots,O_N\}$,
 is straightforward. The basis vectors of $\HH_{\rm phys}$ read 
 $\ket{\vec n}$ = $\ket{n_1,\cdots,n_N}$ where $n_k$ is the charge 
 (or winding number) of a loop encircling the $k^{\rm th}$ singular 
 point, all the other singular points remaining outside of it. 
 The corresponding wave functional is explicitly given by
\eq 
\vev{A|\vec n}= \Psi_{\vec n}[A]=\exp(i\sum_{k=1}^N n_kQ_k)\,,
\eqn{basis-N-punctures}
where $Q_k$ is the flux associated to the $k^{\rm th}$ singular point, defined by:
\eq 
\exp(iQ_k) = h_{C_k}[A]\,, 
\eqn{flux_k}
where 
\eq\ba{l}
C_k = \mbox{closed loop encircling positively one time the singular 
point}\ O_k \\ 
\phantom{C_k = }\mbox{ and leaving aside all the other ones.} 
\ea\eqn{Ck}
The orthonormality relations are
\[
\vev{n_1,\cdots,n_N|n'_1,\cdots,n'_{N'}} = 
\d_{NN'}\prod_{k=1}^N\d_{n_k n_k'}\,.
\]
$\HH_{\rm phys}$ is separable.

One remarks that diffeomorphism invariance, which in the classical theory is a consequence of its gauge invariances, is explicit in the quantum theory constructed here, once all constraints are fulfilled. Note that the states of the (non-separable) kinematical Hilbert space, which still do not obey  the curvature constraint $\CC_2$, are not diffeomorphism invariant since they depend on the location and form of the associated graphs.

\subsection{Observables}
It follows from the above discussion that no non-trivial observables do exist in the case of a trivial topology such as that of $\mathbb{R}^2$. On the other side,  with a non-trivial topology such as that of $\mathbb{R}^2$ with $N$ singular points $O_k$, there is  a a set of $N$ observables $\hat L_k$, $k=1,\cdots,N$, simultaneously diagonalized in the basis \equ{basis-N-punctures} of  $\HH_{\rm phys}$:
\eq 
\hat L_k\ket{\vec n} = n_k\ket{\vec n}\,,\quad k=1,\cdots,N\,.
\eqn{eigen-eq}
They are explicitly given by
\[
\hat L_k = \dint_\S d^2x\,  X^{(k)}_a(\bsx)\hat B^a(\bsx)\,,
\]
where $X^{(k)}_a$ is a closed 1-form ($dX^{(k)}=0$), such that its 
integral on a loop $C_k$ as defined by \equ{Ck}, 
takes the value $i/\h$, whereas its integral on a loop $C_l$ around another singular point $O_l$ vanishes.  Explicitly:
\eq
\dint_{C_k}  X^{(l)} = \dfrac{i}{\h} \d_{kl}\,,
\eqn{intC}
the result depending only on the homotopy class of $C_k$.
In a polar coordinate frame $(r,\theta)$ centred in $O_k$, a particular
solution\footnote{A ``physical'' interpretation may be to view $-iX$ as a 2-dimensional magnetic field whose source is a point current of 
magnitude $1/\h$ located in $O_k$.} 
for the 1-form $X^{(k)}$ is given by  $(A_r=0$ and 
$A_\theta=i/(2\pi\h)$.
The result \equ{eigen-eq} follows from the expression \equ{basis-N-punctures} for 
the basis vector functionals, together with \equ{flux_k} and the differentiation formula (taking into account the support property of  $X^{(k)}$)
\[
\dint_\S d^2x\, X^{(k)}_a(\bsx)\dfud{}{A_a(\bsx)} h_{C_k}[A]
= \lp\dint_{C_k} X\rp h_{C_k}[A]\,.
\]
The operators $\hat L_k$ thus defined are obviously self-adjoint 
in $\HH_{\rm phys}$, and form a complete commutative set of observables.
 
\section{Conclusions}

What we have shown, using the Dirac canonical scheme together with 
the LQG quantization procedure, is that the three-dimensional 
Abelian BF model minimally coupled to a  scalar field obeying a
$\s$-model type of constraint, has the same degrees of 
freedom as the pure BF model. These degrees of freedom 
are non-local, of purely topological nature, characterized 
by the topological nature of space.
They are represented  by a complete set 
of $N$ commuting  observables $\hat L_k$ in the case of the 
space topology being that o $\mathbb{R}^2$ with $N$ points 
ommitted ($N$ ``punctures'').

The generalization to a non-Abelian version is not straightforward 
and will be presented in a future work~\cite{Diego-Olivier}.


\end{document}